\documentclass[aps,showpacs,nofootinbib,superscriptaddress]{revtex4}

\usepackage{amsmath}
\usepackage{graphicx}
\usepackage{float}
\usepackage{bm}
\usepackage{url}

\begin{document}

\title{Boer-Mulders effect in the unpolarized pion induced Drell-Yan process at COMPASS within TMD factorization}
\author{Xiaoyu Wang}
\affiliation{School of Physics, Southeast University, Nanjing 211189, China}
\author{Wenjuan Mao}
\affiliation{School of Physics and Telecommunication Engineering,
Zhoukou Normal University,
Zhoukou 466000, China
}
\author{Zhun Lu}
\email{zhunlu@seu.edu.cn}
\affiliation{School of Physics, Southeast University, Nanjing 211189, China}

\begin{abstract}

We investigate the theoretical framework of the $\cos 2\phi$ azimuthal asymmetry contributed by the coupling of two Boer-Mulders functions in the dilepton production unpolarized $\pi p$ Drell-Yan process by applying the transverse momentum dependent factorization at leading order. We adopt the model calculation results of the unpolarized distribution function $f_1$ and Boer-Mulders function $h_1^\perp$ of pion meson from the light-cone wave functions. We take into account the transverse momentum evolution effects for both the distribution functions of pion and proton by adopting the existed extraction of the nonperturbative Sudakov form factor for the pion and proton distribution functions. An approximate kernel is included to deal with the energy dependence of the Boer-Mulders function related twist-3 correlation function $T_{q,F}^{(\sigma)}(x,x)$ needed in the calculation. We numerically estimate the Boer-Mulders asymmetry $\nu_{BM}$ as the functions of $x_p$, $x_\pi$, $x_F$ and $q_T$ considering the kinematics at COMPASS Collaboration.

\end{abstract}


\maketitle

\section{Introduction}

The Boer-Mulders function is a transverse momentum dependent (TMD) parton distribution function (PDF) that describes the transverse-polarization asymmetry of quarks inside an
unpolarized hadron~\cite{Boer:1999mm,Boer:1997nt}.
Arising from the correlation between the quark transverse spin and the quark transverse momentum, the Boer-Mulders function manifests novel spin structure of hadrons~\cite{Lu:2016pdp}.
For a while the very existence of the Boer-Mulders function was not as obvious.
This is because, similar to its counterpart, the Sivers function, the Boer-Mulders function was thought to be forbidden by the time-reversal invariance of QCD~\cite{Collins:1992kk}.
For this reason, they are classified as T-odd distributions.
However, model calculations incorporating gluon exchange between the struck
quark and the spectator~\cite{Brodsky:2002cx,Brodsky:2002rv}, together with a re-examination~\cite{Collins:2002kn} on the time-reversal argument, show that T-odd distributions actually do not vanish.
It was found that the gauge-links~\cite{Collins:2002kn,Ji:2002aa,Belitsky:2002sm,Boer:2003cm} in the operator definition of TMD distributions play an essential role for a nonzero Boer-Mulders function.

As a chiral-odd distribution, the Boer-Mulders function has to be coupled with another chiral-odd distribution/fragmentation function to survive in a high energy scattering process.
Two promising processes for accessing the Boer-Mulders function are the Drell-Yan and the semi-inclusive deep inelastic scattering (SIDIS) processes. In the former case, the corresponding observables are the $\cos2\phi$ azimuthal angular dependence of the final-state dilepton, which is originated by the convolution of two Boer-Mulders functions from each hadron.
This effect was originally proposed by Boer~\cite{Boer:1999mm} to explain the violation of the Lam-Tung relation observed in $\pi N$ Drell-Yan process\cite{Lam:1978pu}, a phenomenon which cannot be understood from purely perturbative QCD effects \cite{Chiappetta:1986yg,Brandenburg:1993cj,Falciano:1986wk}.
Similar asymmetry was also observed in the $pd$ and $pp$ Drell-Yan processes, and the corresponding data were applied to extract the proton Boer-Mulders function~\cite{Zhang:2008nu,Lu:2009ip,Barone:2009hw,Barone:2010gk}.
Besides the parmaterizations, the Boer-Mulders function of the proton has also been studied extensively in literature by several QCD inspired quark models, such as the spectator model~\cite{Boer:2002ju,Gamberg:2003ey,Bacchetta:2003rz,Lu:2006ew,Gamberg:2007wm,Burkardt:2007xm,Bacchetta:2008af}, the large $N_c$ model~\cite{Pobylitsa:2003ty}, the bag model~\cite{Yuan:2003wk,Courtoy:2009pc} and the light-front constituent quark model~\cite{Courtoy:2009pc,Pasquini:2010af}.
The study of the Boer-Mulders function has been extended to the case of pion meson by the spectator model~\cite{Lu:2004au,Meissner:2008ay,Gamberg:2009uk}, the light-front constituent quark model~\cite{Pasquini:2014ppa,Wang:2017onm} and the bag model~\cite{Lu:2012hh}.

A suitable theoretical framework for studying the $\cos 2\phi$ asymmetry at low transverse momentum is the TMD factorization.
As the TMD evolution of Boer-Mulders function is difficult to solve,
early phenomenological studies focusing on the Boer-Mulders effect in the $\cos2\phi$ asymmetry Drell-Yan~\cite{Boer:2002ju,lu_04,bianconi_04,Lu:2005rq,bianconi_05a,sissakian_05a,gamberg_05,
sissakian_05b,Lu:2006ew,barone_06,lu_07a,reimer_07,miller_07,bianconi_08,
sissakian_08,Lu:2011mz,Liu:2012fha,Liu:2012vn} usually employed tree-level factorization, in which the full TMD evolution of the Boer-Mulders function was not considered.
In Ref.~\cite{Pasquini:2014ppa}, the authors applied a Gaussian ansatz to estimate the $k_T$-evolution effect of the Boer-Mulders function and the $\cos2\phi$ asymmetry in $\pi p$ Drell-Yan process, following the effective description on the energy-dependent broadening of transverse momentum in Ref.~\cite{Schweitzer:2010tt}.
In Ref.~\cite{Wang:2017onm}, the $\cos2\phi$ asymmetry in $\pi p$ Drell-Yan process was studied in a transverse momentum weighted approach. In that work, the weighted asymmetry was expressed as the product of the first $k_T$-moment of the Boer-Mulders function $h_1^{\perp (1)}$, with the scale evolution of $h_1^{\perp (1)}$ included.
In Ref.~\cite{Boer:2001he}, the Collins-Soper-Sterman formalism~\cite{Collins:1981uk,Collins:1984kg,Collins:2011zzd} was applied to study the azimuthal spin asymmetries in electron-positron annihilation which is similar to the case of the Drell-Yan process, and a Sudakov suppression of the asymmetries in the region $q_T \ll Q$ was found.

The purpose of this work is to apply the TMD factorization to estimate the $\cos 2\phi$ azimuthal asymmetry in the pion induced Drell-Yan process contributed by the Boer-Mulders effect. From the viewpoint of TMD factorization~\cite{Collins:1981uk,Collins:1984kg,Collins:2011zzd,Ji:2004xq}, the physical observables can be written as the convolution of the factors related to hard
scattering and well-defined TMD distribution functions or fragmentation functions.
The evolution of TMD functions is usually performed in the $b$ space, which is conjugate to the transverse momentum $\bm{k}_T$~\cite{Collins:1984kg,Collins:2011zzd} through Fourier Transformation.
In the large $b$ region, the $b$ dependence of the TMD distributions and the evolution kernel is nonperturbative.
While in the small $b$ region~(perturbative region), the perturbative methods can be employed and the TMD distributions at fixed energy scale can be expressed as the convolution of perturbatively calculable coefficients $C$ and their collinear counterparts order by order of the $\alpha_s$. The collinear counterparts can be the corresponding collinear parton distribution functions, fragmentation functions or multiparton correlation functions. Particularly, in the case of the Boer-Mulders function, it can be written as the convolution of the perturbatively calculable coefficients and the twist-3 chiral-odd correlation function $T^{(\sigma)}_{q,F}$. The energy dependence of $T_{q,F}^{(\sigma)}(x,x)$ needed in the work can be solved by considering an approximate evolution kernel.

After solving the evolution equations, the TMD evolution from one energy scale to another energy scale is implemented by the exponential factor of the so-called Sudakov-like form factors~\cite{Collins:1984kg,Collins:2011zzd,Collins:1999dz}. The Sudakov-like form factor can be separated into the perturbatively calculable part $S_\mathrm{P}$ and the nonperturbative part $S_{\mathrm{NP}}$, which cannot be perturbatively calculated and only can be extracted from the experiment data.
In Ref.~\cite{Wang:2017zym}, the nonperturbative Sudakov form
factor for the pion distribution functions was extracted from the unpolarized $\pi N$ Drell-Yan data measured by the E615 experiment~\cite{Conway:1989fs} at Fermi Lab. As for the $S_\mathrm{NP}$ related to proton distribution functions, there were several parameterizations~\cite{Landry:2002ix,Su:2014wpa,Aidala:2014hva,Echevarria:2014xaa}.
In this work, we will use the extracted $S_\mathrm{NP}$ of pion for both the evolution of the unpolarized distribution function and the Boer-Mulders function.
As for the evolution of the proton TMD distributions, we will apply the extracted $S_\mathrm{NP}$ in Ref.~\cite{Su:2014wpa} and the parametrization results of the Boer-Mulders function in Ref.~\cite{Lu:2009ip}.

Since the pion meson can serve as the beam to collide off the nucleon target in experiments, the Drell-Yan process~\cite{Drell:1970wh,Drell:1970yt} may be an ideal way to study the parton structure of unstable particles like pions. The idea was brought out decades ago and was exploited by the NA10 Collaboration~\cite{na10} and the E615 Collaboration~\cite{conway}, which measured the azimuthal angular asymmetries in the process $\pi^-\, N\rightarrow \mu^+\mu^- \,X$, with $N$ denoting a nucleon in the deuterium or tungsten target.
Recently, COMPASS Collaboration at CERN~\cite{Gautheron:2010wva,Aghasyan:2017jop,Adolph:2016dvl} started a new Drell-Yan program by colliding a $\pi^-$ meson with energy $E_\pi=190\mathrm{GeV}$ on the $\mathrm{NH}_3$ target, which can be a great opportunity to explore the Boer-Mulders function of the pion meson as well as the nucleon, in the case an unpolarized target or averaging the polarized data can be applied.

The rest of the paper is organized as follows. In Sec.~\ref{Sec.evolution}, we investigate the TMD evolution of the unpolarized distribution function and the Boer-Mulders function of proton and pion meson. In Sec.~\ref{Sec.formalism}, we present the theoretical framework of the $\cos 2\phi$ azimuthal asymmetry $\nu_{BM}$ contributed by the coupling of two Boer-Mulders functions in the pion induced unpolarized Drell-Yan process under the TMD factorization framework. We make the numerical estimate of the $\cos 2\phi$ azimuthal asymmetry in Sec.~\ref{Sec.numerical} and summarize this work in Sec.~\ref{Sec.conclusion}.

\section{The TMD evolution of the distribution functions}
\label{Sec.evolution}

In this section, we will present the TMD evolution formalism of both the unpolarized distribution function $f_1$ and the Boer-Mulders function $h_1^{\perp}$ of the pion as well as those of the proton, within the TMD factorization.
In general, it is more convenient to solve the evolution equations for the TMD distributions in the coordinate space~($\bm{b}$ space) other than that in the transverse momentum $\bm{k}_T$ space, where $\bm{b}$ is conjugate to $\bm{k}_T$ via Fourier transformation~\cite{Collins:1984kg,Collins:2011zzd}.
The TMD distribution functions $\tilde{F}(x,b;\mu,\zeta_F)$ in the $\bm{b}$ space have two energy dependencies, namely, $\mu$ is the renormalization scale related to the corresponding collinear PDFs, and $\zeta_F$ is the energy scale serving as a cutoff to
regularize the light-cone singularity in the operator definition of the TMD distributions.
Here, $F$ is a shorthand for any TMD distribution function and the tilde denotes that the distribution is the one in the $\bm{b}$ space.
The energy evolution for the $\zeta_F$ dependence of the TMD distributions is encoded in the Collins-Soper~(CS)~\cite{Collins:2011zzd} equation:
\begin{align}
\frac{\partial\ \mathrm{ln} \tilde{F}(x,b;\mu,\zeta_F)}{\partial\ \sqrt{\zeta_F}}=\tilde{K}(b;\mu),
\end{align}
while the $\mu$ dependence is derived from the renormalization group equation as
\begin{align}
&\frac{d\ \tilde{K}}{d\ \mathrm{ln}\mu}=-\gamma_K(\alpha_s(\mu)),\\
&\frac{d\ \mathrm{ln} \tilde{F}(x,b;\mu,\zeta_F)}
{d\ \mathrm{ln}\mu}=\gamma_F(\alpha_s(\mu);{\frac{\zeta^2_F}{\mu^2}}),
\end{align}
with $\tilde{K}$ the CS evolution kernel, and $\gamma_K$ and $\gamma_F$ the anomalous dimensions.
The overall structure of the solution for $\tilde{F}(x,b;\mu,\zeta_F)$ is the same as that for the
Sudakov form factor. More specifically, the energy evolution of TMD distributions from an  initial energy $\mu$ to another energy $Q$ is encoded in the Sudakov-like form factor $S$ by the exponential form $\mathrm{exp}(-S)$
\begin{equation}
\tilde{F}(x,b,Q)=\mathcal{F}\times e^{-S}\times \tilde{F}(x,b,\mu),
\label{eq:f}
\end{equation}
where $\mathcal{F}$ is the factor related to the hard scattering. Hereafter, we will set $\mu=\sqrt{\zeta_F}=Q$ and express $\tilde{F}(x,b;\mu=Q,\zeta_F=Q^2)$ as $\tilde{F}(x,b;Q)$ for simplicity.

Studying the $b$-dependence of the TMD distributions can provide useful information regarding the transverse momentum dependence of the hadronic 3D structure through Fourier transformation, which makes the understanding of the $b$ dependence quite important. In the small $b$ region, the $b$ dependence is perturbatively calculable, while in the large $b$ region, the dependence turns to nonperturbative and should be obtained from the experimental data.
To combine the perturbative information at small $b$ with the nonperturbative part at large $b$, a matching procedure must be introduced with a parameter $b_{\mathrm{max}}$ serving as the boundary between the two regions.
A $b$-dependent function $b_\ast$ is defined to have the property $b_\ast\approx b$ at low values of $b$ and $b_{\ast}\approx b_{\mathrm{max}}$ at large $b$ values.
The typical value of $b_{\mathrm{max}}$ is chosen around $1\ \mathrm{GeV}^{-1}$ to guarantee that $b_{\ast}$ is always in the perturbative region.
There are several different $b_\ast$ prescriptions in literature~\cite{Collins:2016hqq,Bacchetta:2017gcc}. In this work we adopt the original prescription introduced in Ref.~\cite{Collins:1984kg} as $b_{\ast}=b/\sqrt{1+b^2/b^2_{\mathrm{max}}}$.

In the small $b$ region $1/Q \ll b \ll 1/ \Lambda$, the TMD distributions at fixed energy $\mu$ can be expressed as the convolution of the perturbatively calculable hard coefficients and the corresponding collinear counterparts, which could be the collinear PDFs or the multiparton correlation functions~\cite{Collins:1981uk,Bacchetta:2013pqa}
\begin{equation}
\tilde{F}_{q/H}(x,b;\mu)=\sum_i C_{q\leftarrow i}\otimes F_{i/H}(x,\mu),
\label{eq:small_b_F}
\end{equation}
where $\otimes$ stands for the convolution in the momentum fraction $x$
\begin{equation}
 C_{q\leftarrow i}\otimes F_{i/H}(x,\mu)\equiv \int_{x}^1\frac{d\xi}{\xi} C_{q\leftarrow i}(x/\xi,b;\mu)F_{i/H}(\xi,\mu)
 \label{eq:otimes}
\end{equation}
and $F_{i/H}(\xi,\mu)$ is the corresponding collinear counterpart of flavor $i$ in hadron $H$ at the energy scale $\mu$, which could be a dynamic scale related to $b_*$ by $\mu_b=c_0/b_*$, with $c_0=2e^{-\gamma_E}$ and the Euler Constant $\gamma_E\approx0.577$~\cite{Collins:1981uk}.

The Sudakov-like form factor $S$ in Eq.~(\ref{eq:f}) can be separated into the perturbatively calculable part $S_{\mathrm{P}}$ and the nonperturbative part $S_{\mathrm{NP}}$
\begin{equation}
\label{eq:S}
S=S_{\mathrm{P}}+S_{\mathrm{NP}}.
\end{equation}
According to the studies in Refs.~\cite{Echevarria:2014xaa,Kang:2011mr,Aybat:2011ge,Echevarria:2012pw,Echevarria:2014rua},
the perturbative part of the Sudakov form factor $S_{P}$ has the same result in Eq.~(\ref{eq:Spert}) among different kinds of distribution functions, i.e., $S_{P}$ is spin-independent.
The perturbative part has the form
\begin{equation}
\label{eq:Spert}
S_{\mathrm{P}}(Q,b_\ast)=\int^{Q^2}_{\mu_b^2}\frac{d\bar{\mu}^2}{\bar{\mu}^2}\left[A(\alpha_s(\bar{\mu}))
\mathrm{ln}\frac{Q^2}{\bar{\mu}^2}+B(\alpha_s(\bar{\mu}))\right].
\end{equation}
The coefficients $A$ and $B$ in Eq.(\ref{eq:Spert}) can be expanded as the series of $\alpha_s/{\pi}$:
\begin{align}
A=\sum_{n=1}^{\infty}A^{(n)}(\frac{\alpha_s}{\pi})^n,\\
B=\sum_{n=1}^{\infty}B^{(n)}(\frac{\alpha_s}{\pi})^n.
\end{align}

In this work, we will take $A^{(n)}$ to $A^{(2)}$ and $B^{(n)}$ to $B^{(1)}$ in the accuracy of next-to-leading-logarithmic (NLL) order~\cite{Collins:1984kg,Landry:2002ix,Qiu:2000ga,Kang:2011mr,Aybat:2011zv,Echevarria:2012pw} :
\begin{align}
A^{(1)}&=C_F,\\
A^{(2)}&=\frac{C_F}{2}\left[C_A\left(\frac{67}{18}-\frac{\pi^2}{6}\right)-\frac{10}{9}T_Rn_f\right],\\
B^{(1)}&=-\frac{3}{2}C_F.
\end{align}

The values of the strong coupling $\alpha_s(\mu)$ are obtained at 2-loop order as an approximation
\begin{align}
\alpha_s(\mu)&=\frac{12\pi}{(33-2n_f)\mathrm{ln}(\mu^2/\Lambda^2_{QCD})}
\left\{{1-\frac{6(153-19n_f)}{(33-2n_f)^2}
\frac{\mathrm{ln}\mathrm{ln}(\mu^2/\Lambda^2_{QCD})}{\mathrm{ln}(\mu^2/\Lambda^2_{QCD})}}\right\}, \label{eq:alphas}
\end{align}
with fixed $n_f=5$ and $\Lambda_{\mathrm{QCD}}=0.225\ \mathrm{GeV}$. We note that the running coupling in Eq.~(\ref{eq:alphas}) satisfies $\alpha_s(M_Z^2)=0.118$.
The quark and antiquark contributes to the perturbative part $S_{\mathrm{P}}$ equally~\cite{Prokudin:2015ysa}, i.e.,
\begin{align}
S^q_{\mathrm{P}}(Q,b_\ast)=S^{\bar{q}}_{\mathrm{P}}(Q,b_\ast)=S_{\mathrm{P}}(Q,b_\ast)/2.
\end{align}

For the nonperturbative form factor $S_{\mathrm{NP}}$ associated with the unpolarized distribution of the proton, a general parameterization has been proposed in Ref.~\cite{Su:2014wpa} and it has the form
\begin{align}
S_{\mathrm{NP}}=g_1b^2+g_2\mathrm{ln}\frac{b}{b_{\ast}}\mathrm{ln}\frac{Q}{Q_0}
+g_3b^2\left((x_0/x_1)^{\lambda}+(x_0/x_2)^\lambda\right).
\label{eq:SNP_DY_NN}
\end{align}
In Ref.~\cite{Su:2014wpa} the parameters $g_1,\ g_2,\ g_3$ are fitted from the nucleon-nucleon Drell-Yan process data
at the initial scale $Q^2_0=2.4\ \mathrm{GeV}^2$ with $b_{\mathrm{max}}=1.5\ \mathrm{GeV}^{-1}$, $x_0=0.01$ and $\lambda=0.2$. The parameters are extracted as $g_1=0.212,\ g_2=0.84, \ g_3 = 0$.
Since the nonperturbative form factor $S_{\mathrm{NP}}$ for quarks and antiquarks satisfies the following relation~\cite{Prokudin:2015ysa}
\begin{align}
S^q_{\mathrm{NP}}(Q,b)+S^{\bar{q}}_{\mathrm{NP}}(Q,b)=S_{\mathrm{NP}}(Q,b),
\end{align}
$S_{\mathrm{NP}}$ associated with the TMD distribution function for the protons can be expressed as
\begin{align}
\label{eq:SNPproton}
S^{f_{1,q/p}}_{\mathrm{NP}}(Q,b)=\frac{g_1}{2}b^2+\frac{g_2}{2}\ln\frac{b}{b_{\ast}}\ln\frac{Q}{Q_0}.
\end{align}
In this work, we will apply the above result to calculate the spin-independent cross-section.

In Ref.~\cite{Wang:2017zym}, parameterization of the nonperturbative Sudakov form factor for the unpolarized TMD distribution of the pion was proposed as follows:
\begin{align}
S^{f_{1,q/\pi}}_{\mathrm{NP}}=g^\pi_1b^2+g^\pi_2\mathrm{ln}\frac{b}{b_{\ast}}\mathrm{ln}\frac{Q}{Q_0},
\label{eq:SNP_pion}
\end{align}
which has the same form as that for the proton.
Here the parameters $g_1^\pi$ and $g_2^\pi$ were fitted at the initial energy scale $Q^2_0=2.4\ \mathrm{GeV}^2$ with $b_{\mathrm{max}}=1.5\ \mathrm{GeV}^{-1}$ as $g^\pi_1=0.082$ and $g^\pi_2=0.394$.
We note that a form of $S^{f_{1,q/\pi}}_{\mathrm{NP}}$ motivated by the NJL model was given in Ref.~\cite{Ceccopieri:2018nop}.

Thus we can rewrite the scale-dependent TMD distribution function $\tilde{F}$ of the proton and the pion in $b$ space as
\begin{align}
\label{eq:tildeF}
\tilde{F}_{q/p}(x,b;Q)=e^{-\frac{1}{2}S_{\mathrm{P}}(Q,b_\ast)-S^{F_{q/p}}_{\mathrm{NP}}(Q,b)}
\mathcal{F}(\alpha_s(Q))\sum_i C_{q\leftarrow i}\otimes F_{i/p}(x,\mu_b),\\
\tilde{F}_{q/\pi}(x,b;Q)=e^{-\frac{1}{2}S_{\mathrm{P}}(Q,b_\ast)-S^{F_{q/\pi}}_{\mathrm{NP}}(Q,b)}
\mathcal{F}(\alpha_s(Q))\sum_i C_{q\leftarrow i}\otimes F_{i/\pi}(x,\mu_b).
\end{align}


The hard coefficients $C_i$ and $\mathcal{F}$ for $f_1$ have been calculated up to next-to-leading order (NLO), while those for the Boer-Mulders function are still remained in leading order (LO).
For consistency, in this work we will adopt the LO results of the $C$ coefficients for  $f_{1,q/H}$ and $h_{1,q/H}^{\perp}$, i.e., we take $C_{q\leftarrow i}=\delta_{qi}\delta(1-x)$ and take the hard factor $\mathcal{F} =1$.

Thus, we can obtain the unpolarized distribution function of the proton and pion in $b$ space as
\begin{align}
\tilde{f}_{1,q/p}(x,b;Q) &=e^{-\frac{1}{2}S_{\mathrm{P}}(Q,b_\ast)-S^{f_{1, q/p}}_{\mathrm{NP}}(Q,b)}
 f_{1, q/p}(x,\mu_b),\nonumber\\
\tilde{f}_{1,q/\pi}(x,b;Q) &=e^{-\frac{1}{2}S_{\mathrm{P}}(Q,b_\ast)-S^{f_{1, q/\pi}}_{\mathrm{NP}}(Q,b)}
 f_{1, q/\pi}(x,\mu_b).
\label{eq:f_b}
\end{align}

If we perform a Fourier transformation on the $\tilde{f}_{1,q/H}(x,b;Q)$, we can obtain the distribution function in the transverse momentum space as
\begin{align}
f_{1,q/p}(x,k_T;Q)=\int_0^\infty\frac{dbb}{2\pi}J_0(k_T b)e^{-\frac{1}{2}S_{\mathrm{P}}(Q,b_\ast)-S^{f_{1, q/p}}_{\mathrm{NP}}(Q,b)} f_{1, q/p}(x,\mu_b),\\
f_{1,q/\pi}(x,k_T;Q)=\int_0^\infty\frac{dbb}{2\pi}J_0(k_T b)e^{-\frac{1}{2}S_{\mathrm{P}}(Q,b_\ast)-S^{f_{1, q/\pi}}_{\mathrm{NP}}(Q,b)} f_{1, q/\pi}(x,\mu_b),
\end{align}
where $J_0$ is the Bessel function of the first kind, and $k_T = |\bm k_T|$.

According to Eq.~(\ref{eq:small_b_F}), in the small $b$ region, we can also express the Boer-Mulders at one fixed energy scale in terms of the perturbatively calculable coefficients and the corresponding collinear correlation function
\begin{align}
\widetilde{h}_{1,q/H}^{\alpha\perp}(x,b;\mu)=(\frac{-ib^\alpha}{2})T^{(\sigma)}_{q/H,F}(x,x;\mu),
\end{align}
where the hard coefficients are only calculated up to LO.
Here $T^{(\sigma)}_{q/H,F}(x,x;\mu)$ is the chiral-odd twist-3 quark-gluon-quark correlation function and is related to the first transverse moment of the Boer-Mulders function $h_{1,q/H}^{\perp (1)}$ by:
\begin{align}
T^{(\sigma)}_{q/H,F}(x,x;\mu)=\int d^2 \bm{k}_T\frac{\bm{k}_T^2}{M_H}h_{1,q/H}^\perp(x,\bm{k}_T;\mu)
= 2M_H h_{1,q/H}^{\perp (1)} \label{eq:qsbm}
\end{align}

As for the nonperturbative part of the Sudakov form factor associated with the Boer-Mulders function,
the information still remains unknown. In a practical calculation, we assume that it is the
same as $S_{NP}^{f_{1,q/H}}$, i.e., $S_{NP}^{h_{1,q/p}^\perp}=S_{NP}^{f_{1,q/p}} $ and $S_{NP}^{h_{1,q/\pi}^\perp}=S_{NP}^{f_{1,q/\pi}}$.
Therefore, we can obtain the Boer-Mulders functions of the pion and proton in $b$-space as
\begin{align}
\widetilde{h}_{1,q/p}^{\alpha\perp}(x,b;Q)=(\frac{-ib^\alpha}{2})
e^{-\frac{1}{2}S_{\mathrm{P}}(Q,b_\ast)-S^{f_{1 ,q/p}}_{\mathrm{NP}}(Q,b)}
T^{(\sigma)}_{q/p,F}(x,x;\mu_b),\label{eq:protonBM_b}\\
\widetilde{h}_{1,q/\pi}^{\alpha\perp}(x,b;Q)=(\frac{-ib^\alpha}{2})
e^{-\frac{1}{2}S_{\mathrm{P}}(Q,b_\ast)-S^{f_{1,q/\pi}}_{\mathrm{NP}}(Q,b)}
T^{(\sigma)}_{q/\pi,F}(x,x;\mu_b).
\label{eq:BM_b}
\end{align}
After performing the Fourier transformation back to the transverse momentum space, one can get the Boer-Mulders function as
\begin{align}
\frac{k_T}{M_p}h^\perp_{1,q/p}(x,k_T;Q)=\int_0^\infty db(\frac{b^2}{2\pi})J_1(k_T b)e^{-\frac{1}{2}S_{\mathrm{P}}(Q,b_\ast)-S^{f_{1, q/p}}_{\mathrm{NP}}(Q,b)}
h^{\perp(1)}_{1,q/p}(x;\mu_b),\\
\frac{k_T}{M_\pi}h^\perp_{1,q/\pi}(x,k_T;Q)=\int_0^\infty db(\frac{b^2}{2\pi})J_1(k_T b)e^{-\frac{1}{2}S_{\mathrm{P}}(Q,b_\ast)-S^{f_{1, q/\pi}}_{\mathrm{NP}}(Q,b)}
h^{\perp(1)}_{1,q/\pi}(x;\mu_b).
\label{eq:BM_kt}
\end{align}

\section{The $\cos 2\phi$ azimuthal asymmetry contributed by the Boer-Mulders functions in Drell-Yan process}

\label{Sec.formalism}

In this section, by applying the TMD factorization with evolution effect, we set up the necessary framework of the $\cos 2\phi$ azimuthal angular asymmetry contributed by the Boer-Mulders functions in the pion induced unpolarized Drell-Yan process.
In the studied process, the $\pi^-$ beam is scattered off the unpolarized proton target, where the quark and antiquark in the beam and target annihilate into a photon and the photon then produces a lepton pair in the final state. The process can be written as
\begin{align}
\pi^-(P_\pi)+p(P_p)\longrightarrow \gamma^*(q)+X \longrightarrow l^+(\ell)+l^-(\ell')+X,
\end{align}
where $P_\pi$, $P_p$ and $q$ denote the momenta of the $\pi^-$ meson, the proton and the virtual photon, respectively. Here, $q$ is a timelike vector in Drell-Yan process, namely, $Q^2=q^2>0$, which can be interpreted as the invariant mass square of the lepton pair. In order to express the experimental observables, we adopt the following kinematical variables~\cite{Collins:1984kg,Gautheron:2010wva}
\begin{align}
&s=(P_{\pi}+P_p)^2,\quad x_\pi=\frac{Q^2}{2P_\pi\cdot q},\quad x_p=\frac{Q^2}{2P_p\cdot q},\nonumber\\
&x_F=2q_L/s=x_\pi-x_p,\quad\tau=Q^2/s=x_\pi x_p,\quad y=\frac{1}{2}\mathrm{ln}\frac{q^+}{q^-}=\frac{1}{2}\mathrm{ln}\frac{x_\pi}{x_p},
\end{align}
where $s$ is the total center-of-mass~(c.m.) energy squared; $x_\pi$ and $x_p$ are the Bjorken variables; $q_L$ is the longitudinal momentum of the virtual photon in the c.m. frame of the incident hadrons; $x_F$ is the Feynman $x$ variable, which corresponds to the longitudinal momentum fraction carried by the lepton pair; and $y$ is the rapidity of the lepton pair.
In the leading-twist approximation $x_\pi$ and $x_p$ can be interpreted as the momentum fraction carried by the annihilating quark/antiquark inside the $\pi^-$ and the proton, respectively.
Alternatively, $x_\pi$ and $x_p$ can be expressed as functions of $x_F$, $\tau$ and of $y$, $\tau$~\cite{Wang:2017zym}
\begin{align}
x_{\pi/p}=\frac{\pm x_F+\sqrt{x_F^2+4\tau}}{2},\quad x_{\pi/p}=\sqrt{\tau} e^{\pm y}.
\end{align}

The angular differential cross section for unpolarized Drell-Yan process has the following general form~\cite{Lu:2016pdp}
\begin{align}
\frac{1}{\sigma }\frac{d\sigma }{d\Omega }=\frac{3}{4\pi }\frac{1}{\lambda +3}(1+\lambda\cos^2\theta +\mu \sin2\theta \cos\phi+\frac{\nu }{2} \sin^2\theta \cos2\phi),
\label{eq:cross section}
\end{align}
where $\theta$ is the polar angle, and $\phi$ is the azimuthal angle of the hadron plane with respect to the dilepton plane in the Collins-Soper~(CS) frame~\cite{Collins:1977iv}.
The coefficients $\lambda$, $\mu$, $\nu$ in Eq.~(\ref{eq:cross section}) describe the sizes of different angular dependencies.
Particularly, $\nu$ stands for the asymmetry of the $\cos 2\phi$ azimuthal angular distribution of the dilepton.

The coefficients $\lambda$, $\mu$, $\nu$ have been measured in the process $\pi^-\, N\rightarrow \mu^+\mu^- \,X$  by the NA10 Collaboration~\cite{na10} and the E615 Collaboration~\cite{conway} for a $\pi^-$ beam
with energies of 140, 194, 286 GeV~\cite{na10}, and 252 GeV~\cite{conway},
with $N$ denoting a nucleon in the deuterium or tungsten target.
The experimental data showed a large value of $\nu$, near 30\% in the region $Q_T\sim3$ GeV.
This demonstrates a clear violation of the Lam-Tung relation~\cite{Lam:1978pu}.
In the last decade $\lambda$, $\mu$, $\nu$ were also measured in the $p\, d$ and $pp$ Drell-Yan processes~\cite{Zhu:2006gx,Zhu:2008sj}.
The origin of large $\cos2\phi$ asymmetry--or the violation of the Lam-Tung relation-- observed in Drell-Yan process has been studied extensively in literature~\cite{Collins:1978yt,bran93,bran94,Eskola,Boer:1999mm,Boer:2006eq,Blazek,Zhou:2009rp,Peng:2015spa}.
Here we will only consider the contribution from the coupling of the Boer-Mulders functions, denoted by $\nu_{\textrm{BM}}$.
It might be measured through the combination $2\nu_{\textrm{BM}}\approx 2\nu+\lambda-1$, in which the perturbative contribution is largely subtracted.

According to the TMD framework, in the Collins-Soper frame~\cite{Collins:1977iv} the unpolarized Drell-Yan cross section at leading twist can be written as~\cite{Boer:1999mm}
\begin{align}
\frac{d\sigma ({h}_{1}{h}_{2}\rightarrow l\bar{l}X)}{d\Omega d{x}_\pi d{x}_p{d}^{2}{\bm{q}}_{T}}=\frac{{\alpha }^{2}}{3{Q}^{2}}\sum _{q}\bigg{\{ }A(y)\mathcal{F}[f_{1,q/\pi}f_{1,\bar{q}/p}] +B(y)\cos2\phi\mathcal{F}[(2\bm{\hat{h}}\cdot \bm{k}_{T} \bm{\hat{h}}\cdot \bm{p}_{T})-(\bm{k}_T\cdot \bm{p}_T)]\frac{h^\perp_{1,q/\pi} h^\perp _{1,\bar{q}/p}}{{M}_\pi{M}_p}\bigg{\}},
\label{eq:dcs}
\end{align}
where we adopt the notation
\begin{align}
\mathcal{F}[\omega f\bar{f}]= e_q^2\int {d}^{2}\bm{k}_T{d}^{2}\bm{p}_T{\delta
}^{2}(\bm{k}_T+\bm{p}_T-\bm{{q}}_{T})\omega f({x}_\pi,\bm{{k}}_T^2)\bar{f}({x}_p,\bm{p}_T^2)
\label{eq:notation}
\end{align}
to express the convolution of transverse momenta.
Here $\bm{q}_T$, $\bm{k}_T$ and $\bm{p}_T$ are the transverse momenta of the lepton pair, quark and antiquark in the initial hadrons, respectively. $\bm{\hat{h}}$ is a unit vector defined as $\bm{\hat{h}}=\frac{\bm{q}_{T}}{|\bm{q}_{T}|} = \frac{\bm{q}_{T}}{q_T}$.
The second term in Eq.~(\ref{eq:dcs}) has a $\cos 2\phi$ modulation and can contribute to $\nu$ asymmetry.
Two coefficients $A(y)$ and $B(y)$ in Eq.~(\ref{eq:dcs}) can be written as the function of $\theta$ in the c.m. frame of the lepton pair
\begin{align}
&A(y)=(\frac{1}{2}-y+y^2)=\frac{1}{4}(1+\cos^2 \theta),\nonumber\\
&B(y)=y(1-y)=\frac{1}{4} \sin^2 \theta. \nonumber
\end{align}
Combining Eqs.~(\ref{eq:cross section}) and (\ref{eq:dcs}), we can obtain the expression of the $\cos 2\phi$ asymmetry coefficient $\nu_{\textrm{BM}}$ contributed by the Boer-Mulders functions as
\begin{align}
\nu_{\textrm{BM}}&=\frac{2\sum_q\mathcal{F}\left[\left(2\bm{\hat{h}}\cdot \bm{k}_T \bm{\hat{h}}\cdot \bm{p}_T-\bm{k}_{T}\cdot \bm{p}_T\right)\frac{h^\perp _{1,q/\pi}h^\perp_{1,\bar{q}/p}}{M_\pi M_p}\right]}{\sum_q\mathcal{F}\left[f_{1,q/\pi}f_{1,\bar{q}/p}\right]
} .
\label{eq:nu}
\end{align}

Adopting the notation in Eq.~(\ref{eq:notation}) and performing the Fourier transformation from the $\bm{q}_T$ space to $\bm{b}$ space on the delta function, we can obtain the denominator in Eq.~(\ref{eq:nu}) as
\begin{align}
\mathcal{F}\left[f_{1,q/\pi}f_{1,\bar{q}/p}\right]&=\sum_q e_q^2\int \frac{d^2b}{(2\pi)^2} \int d^2\bm{k}_Td^2\bm{p}_T e^{i(\bm{{q}}_{T}-\bm{k}_T-\bm{p}_T)\cdot \bm{b}}f_{1,q/\pi}({x}_\pi,\bm{{k}}_T^2)f_{1,\bar{q}/p}({x}_p,\bm{p}_T^2)\nonumber\\
&=\sum_q e_q^2\int \frac{d^2b}{(2\pi)^2} e^{i\bm{q}_{T}\cdot \bm{b}} \widetilde{f}_{1,q/\pi}(x_\pi,b;Q)\widetilde{f}_{1,\bar{q}/p}(x_p,b;Q)\nonumber\\
&=\sum_q e_q^2\int_0^\infty \frac{b db}{2\pi}J_0(q_Tb)f_{1,q/\pi}(x_\pi,\mu_b)f_{1,\bar{q}/p}(x_p,\mu_b)
e^{-\left(S^{f_{1,q/p}}_{\mathrm{NP}}+S^{f_{1,q/\pi}}_{\mathrm{NP}}+S_\mathrm{P}\right)}
\label{eq:denominator}
\end{align}
where the unpolarized distribution function in $b$ space is given in Eq. (\ref{eq:f_b}).

Similar to the treatment of the denominator, we can write the numerator using the expression of the Boer-Mulders function in Eqs.~(\ref{eq:protonBM_b}) and (\ref{eq:BM_b}) as
\begin{align}
&\mathcal{F}\left[\left(2\bm{\hat{h}}\cdot \bm{k}_T \bm{\hat{h}}\cdot \bm{p}_T-\bm{k}_{T}\cdot \bm{p}_T\right)\frac{h^{\perp}_{1,q/\pi}h^{\perp}_{1,\bar{q}/p}}{M_\pi M_p}\right]\nonumber\\
&=\sum_q e_q^2\int \frac{d^2b}{(2\pi)^2} \int d^2\bm{k}_Td^2\bm{p}_T e^{i(\bm{{q}}_{T}-\bm{k}_T-\bm{p}_T)\cdot b}\left[\left(2\bm{\hat{h}}\cdot \bm{k}_T \bm{\hat{h}}\cdot \bm{p}_T-\bm{k}_{T}\cdot \bm{p}_T\right)\frac{h^{\perp}_{1,q/\pi}h^{\perp}_{1,\bar{q}/p}}{M_\pi M_p}\right] \nonumber\\
&=\sum_q e_q^2\int \frac{d^2b}{(2\pi)^2} e^{i\bm{q}_T\cdot \bm{b}} (2\hat{h}_\alpha\hat{h}_\beta-g_{\alpha\beta}^\perp)\widetilde{h}^{\alpha\perp }_{1,q/\pi}(x_\pi,b;Q)\widetilde{h}^{\beta\perp}_{1,\bar{q}/p}(x_p,b;Q)\nonumber\\
&=\sum_q e_q^2\int \frac{d^2b}{(2\pi)^2} e^{i\bm{q}_T\cdot \bm{b}} (2\hat{h}_\alpha\hat{h}_\beta-g_{\alpha\beta}^\perp)(\frac{-ib^\alpha}{2})
T^{(\sigma)}_{q/\pi,F}(x_\pi,x_\pi;\mu_b)(\frac{-ib^\beta}{2})
T^{(\sigma)}_{\bar{q}/p,F}(x_p,x_p;\mu_b)
e^{-\left(S^{f_{1,q/p}}_{\mathrm{NP}}+S^{f_{1,q/\pi}}_{\mathrm{NP}}+S_\mathrm{P}\right)}\nonumber\\
&=\sum_q e_q^2\int_0^\infty \frac{dbb^3} {8\pi}J_2(q_Tb)T^{(\sigma)}_{q/\pi,F}(x_\pi,x_\pi;\mu_b)
T^{(\sigma)}_{\bar{q}/p,F}(x_p,x_p;\mu_b)
e^{-\left(S^{f_{1,q/p}}_{\mathrm{NP}}+S^{f_{1,q/\pi}}_{\mathrm{NP}}+S_\mathrm{P}\right)}.
\label{eq:numerator}
\end{align}
with $T^{(\sigma)}_{q/\pi,F}(x_\pi,x_\pi;\mu_b)$ and $T^{(\sigma)}_{\bar{q}/p,F}(x_p,x_p;\mu_b)$ the chiral-odd quark-gluon-quark correlation function of the pion and proton defined in Eq.~(\ref{eq:qsbm}).

\section{Numerical estimate}
\label{Sec.numerical}

In this section, using the framework set up above, we present the numerical prediction of the $\cos 2\phi$ azimuthal asymmetry $\nu_{\textrm{BM}}$ in the pion induced unpolarized Drell-Yan process at the kinematics of the COMPASS Collaboration. To do this, we need to know the Boer-Mulders functions of the pion and the proton.
\begin{figure}
  \centering
  \includegraphics[width=0.48\columnwidth]{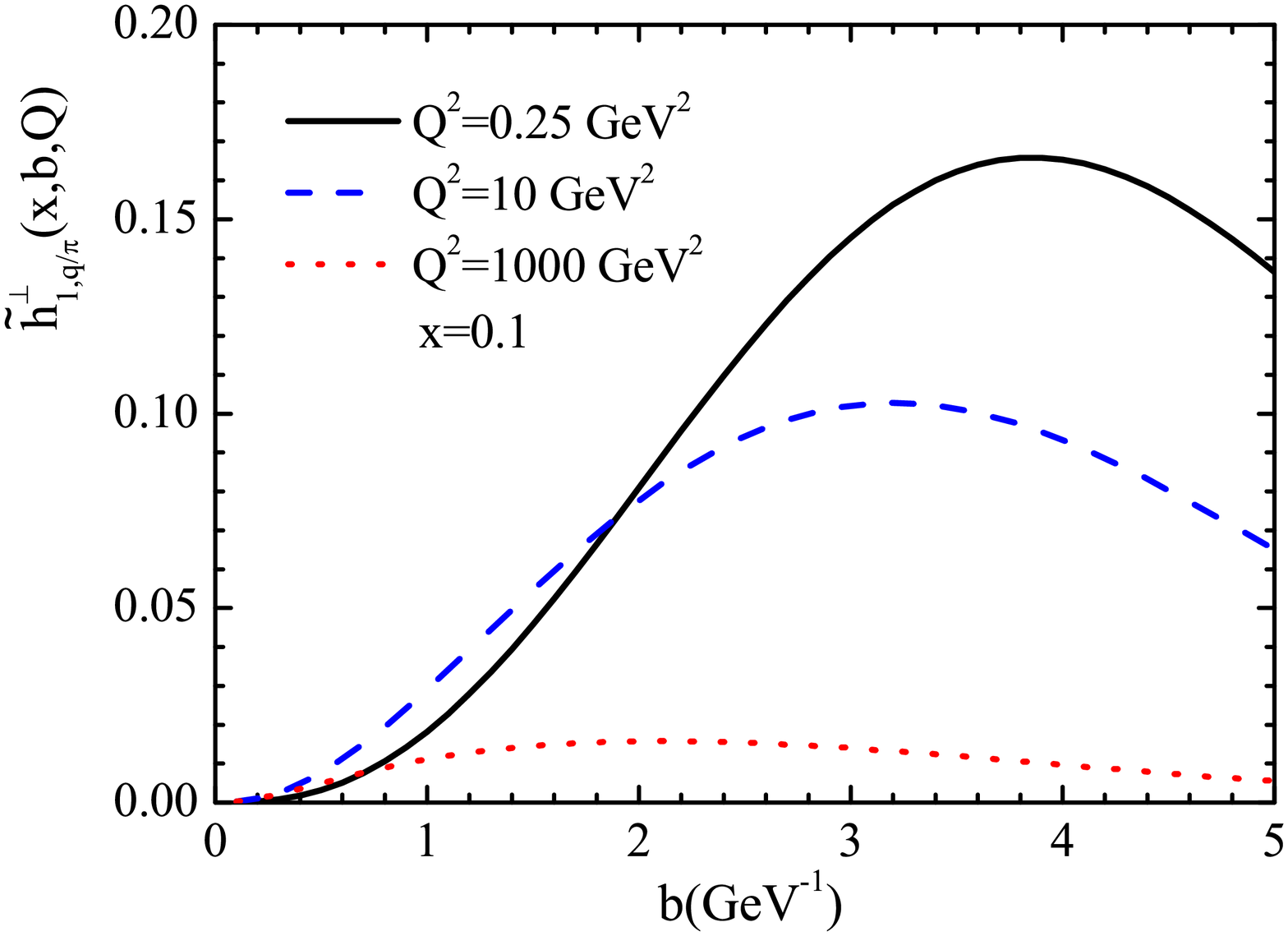}
  \includegraphics[width=0.48\columnwidth]{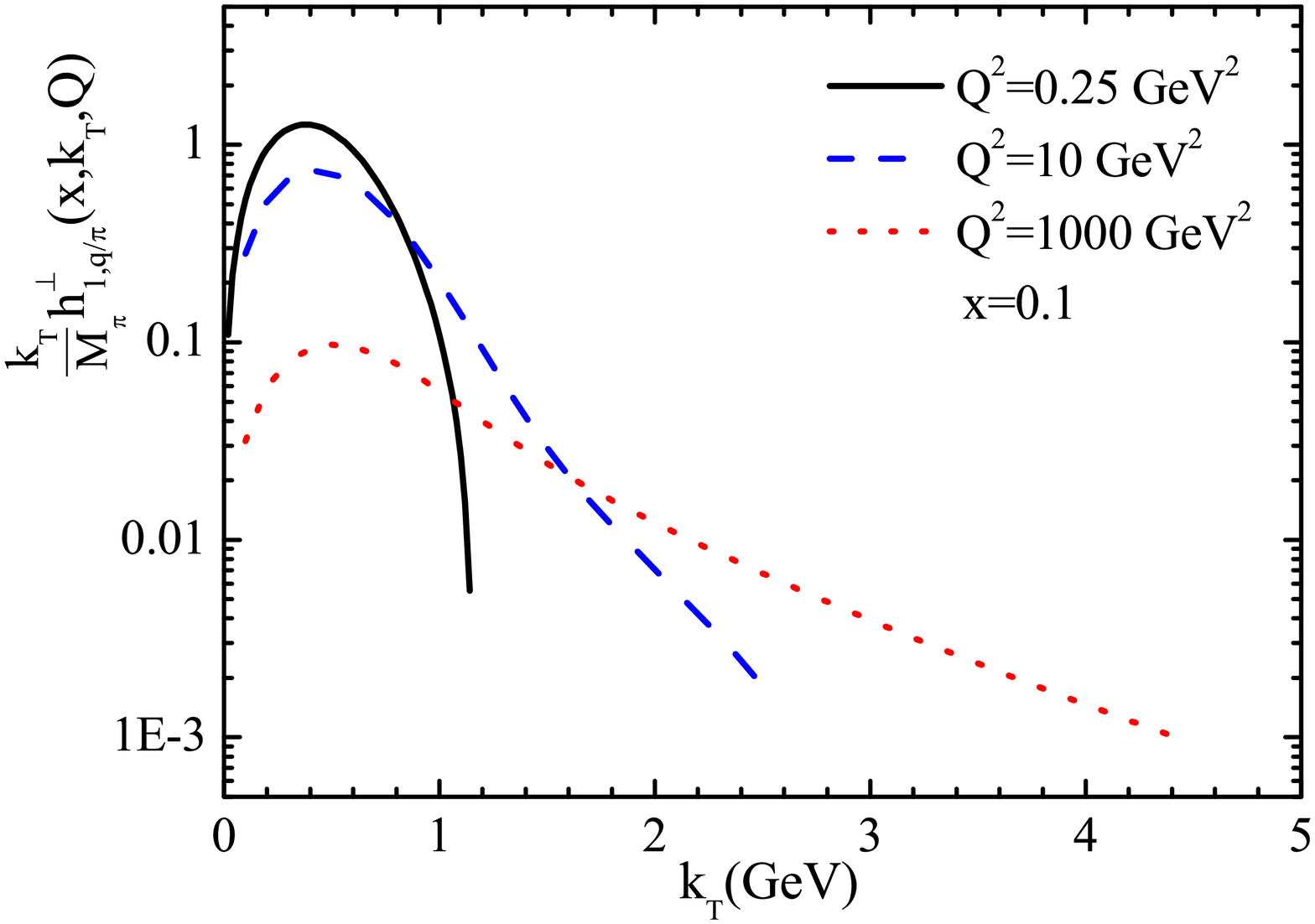}\\
  \caption{The Boer-Mulders function for $u$-quark in $b$ space~(left panel) and $k_T$ space~(right panel) considering three different energy scales: $Q^2=2.4 \mathrm{GeV}^2$~(solid lines), $Q^2=10 \mathrm{GeV}^2$~(dashed lines), $Q^2=1000 \mathrm{GeV}^2$~(dotted lines).}
  \label{fig:BM}
\end{figure}

In Ref.~\cite{Wang:2017onm}, the integrated unpolarized distribution function and the Boer-Mulders function for the pion meson were calculated by a model, in which the pion wave functions is derived from a light-cone approach. In this work we adopt the results at the model scale $\mu_0^2=0.25\mathrm{GeV}^2$ as
\begin{align}
f_{1\pi}(x)&=\frac{A^2}{4\pi^2}\beta^2x(1-x)\mathrm{exp}\left[-\frac{1}{4\beta^2}\frac{m^2}{x(1-x)}\right] ,\nonumber\\
{h}^{\perp }_{1,\pi }(x,\bm{k}^{2}_T)&=\frac{{C}_{F}{\alpha }_{s}}{16{\pi }^{3}}\frac{mM_\pi}{\sqrt{{m}^{2}+\bm{k}^{2}_T}}\frac{{A}^{2}}{\bm{k}^{2}_T}\mathrm{exp}[-\frac{1}{8{\beta }^{2}}\frac{\bm{k}^{2}_T+{m}^{2}}{x(1-x)}]\left[\Gamma (\frac{1}{2},\frac{{m}^{2}}{8\beta^2 x(1-x)})-\Gamma (\frac{1}{2},\frac{\bm{k}^{2}_T+{m}^{2}}{8\beta^2 x(1-x)})\right].
\end{align}
The values of the parameters are as follows~\cite{Xiao:2003wf,Wang:2017onm}
\begin{align}
\beta=0.41\ \textrm{GeV},\quad m_u=m_d=m=0.2\ \textrm{GeV}, \quad A=31.303\ \textrm{GeV}^{-1}.
\end{align}

To perform the evolution of $f_{1,q/\pi}$ from the model scale $\mu_0$ to another energy scale numerically, we apply the {\sc{QCDNUM}} evolution package~\cite{Botje:2010ay}. As for the energy evolution of $T^{(\sigma)}_{q,F}$, the exact evolution effect has been studied in Ref.~\cite{Kang:2012em}.
For our purpose, we only consider the homogenous term in the evolution kernel
\begin{align}
P^{T^{(\sigma)}_{q,F}}_{qq}(x)\approx\Delta_T\,P_{qq}(x)-N_C\delta(1-x),\label{eq:evobm}
\end{align}
with $\Delta_T\,P_{qq}(x)=C_F\left[\frac{2z}{(1-z)_+}+\frac{3}{2}\delta(1-x)\right]$ being the evolution kernel for the transversity distribution function $h_1(x)$. We customize the original code of {\sc{QCDNUM}} to include the approximate kernel in Eq.~(\ref{eq:evobm}).

Applying Eqs.~(\ref{eq:BM_b}) and (\ref{eq:BM_kt}), we calculate the Boer-Mulders function for the up quark inside the $\pi$ meson at different scales.
The results for the $b$ dependent and $k_T$-dependent Boer-Mulders function at $x=0.1$ are plotted in the left and right panels of Fig.~\ref{fig:BM}, respectively.
In calculating $\tilde{h}_{1,q/\pi}^{\perp}(x,b;Q)$ in Fig.~\ref{fig:BM}, we have
rewritten the Boer-Mulders function in $b$ space as
\begin{align}
&\tilde{h}_{1,q/\pi}^{\perp}(x,b;Q)= \frac{i b_\alpha}{\pi} \tilde{h}_{1,q/\pi}^{\alpha\perp}(x,b;Q).
\end{align}
The three curves in each panel correspond to three different energy scales: $Q^2=0.25 \mathrm{GeV}^2$~(solid lines), $Q^2=10 \mathrm{GeV}^2$~(dashed lines), $Q^2=1000 \mathrm{GeV}^2$~(dotted lines).
From the curves, we find that the TMD evolution effect of the Boer-Mulders function is significant and should  be considered in phenomenological analysis.
The result also indicates that the perturbative Sudakov form factor
dominates in the low $b$ region at higher energy scales and the nonperturbative part of the TMD evolution becomes more important at lower energy scales.

For the Boer-Mulders function of the proton needed in the calculation, we adopt the parametrization at the initial energy $Q_0^2=1\mathrm{GeV}^2$ in Ref.~\cite{Lu:2009ip}:
\begin{align}
&{h}^{\perp q}_1(x,\bm{k}^2_T)={h}^{\perp q}_{1}(x)\frac{\exp (-\bm{k}_T^2/k_{bm}^2)}{\pi k_{bm}^2},\\
&{h}^{\perp q}_{1}(x)=H_{q}{x}^{c^{q}}(1-x)^{b}{f}^{q}_{1}(x).
\label{eq:bmproton}
\end{align}
As for the unpolarized distribution function $f_{1,q/p}(x)$ of the proton, we adopt the leading-order set of the MSTW2008 parametrization~\cite{Martin:2009iq}.

\begin{figure}
  \centering
  \includegraphics[width=0.48\columnwidth]{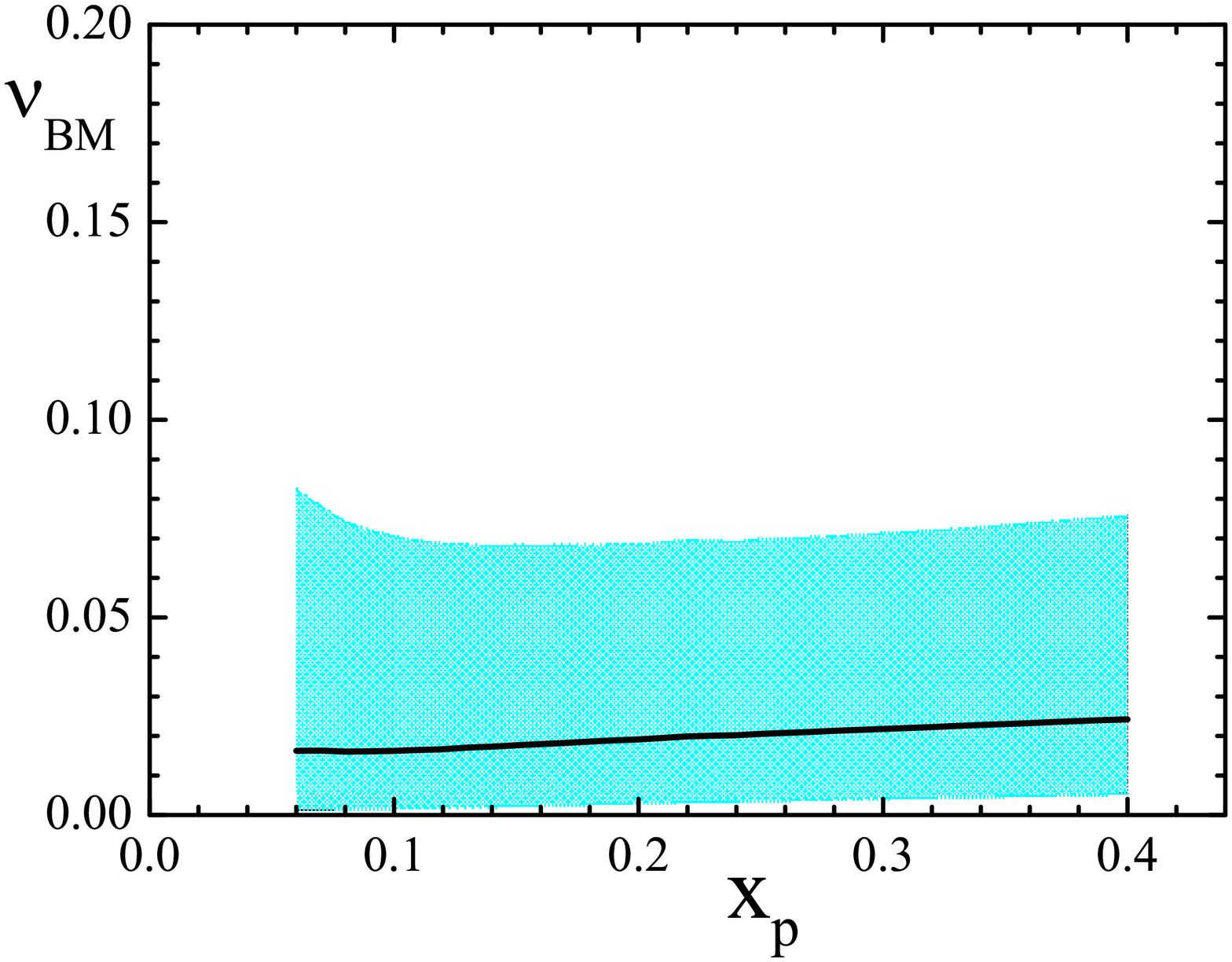}
  \includegraphics[width=0.48\columnwidth]{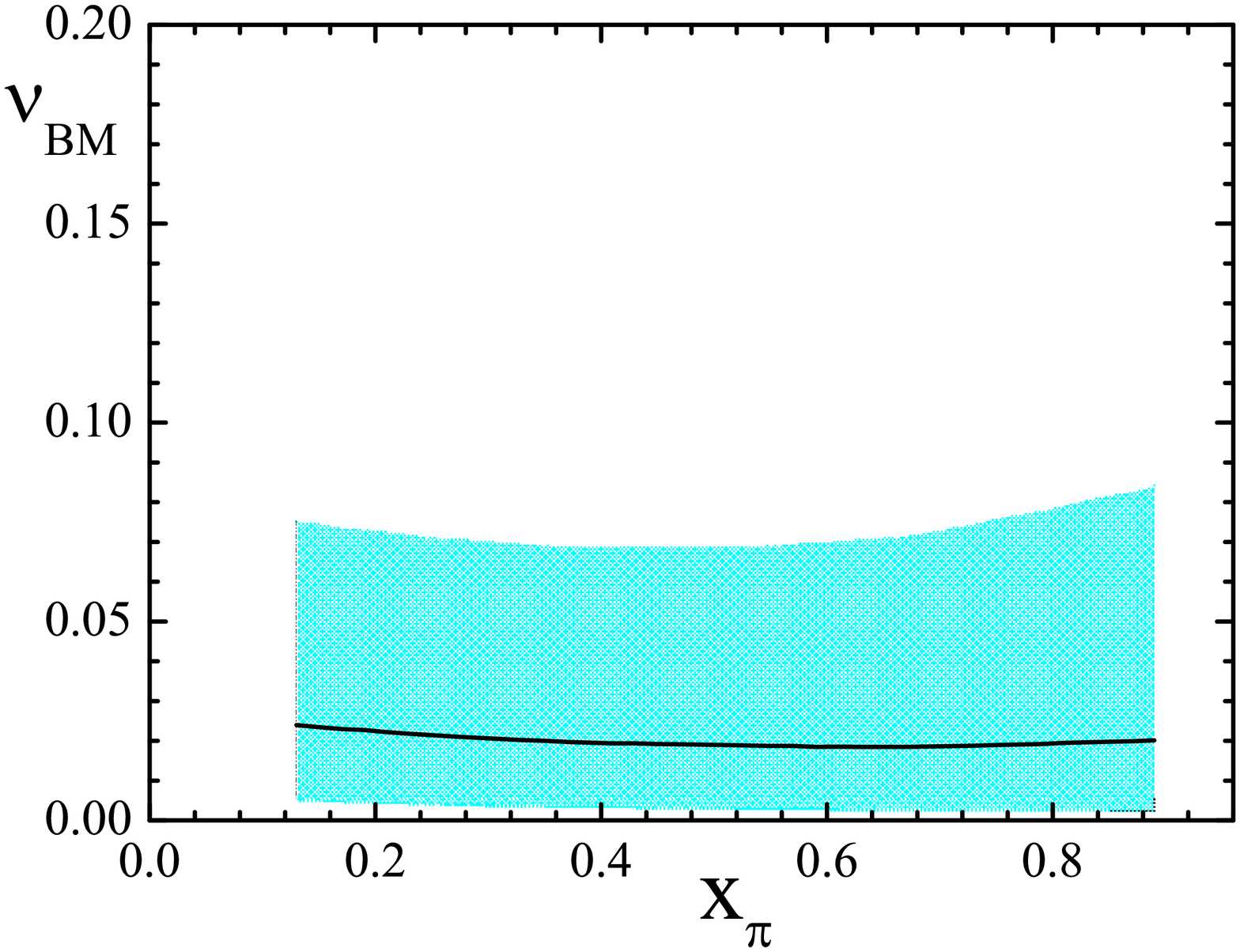}\\
  \includegraphics[width=0.48\columnwidth]{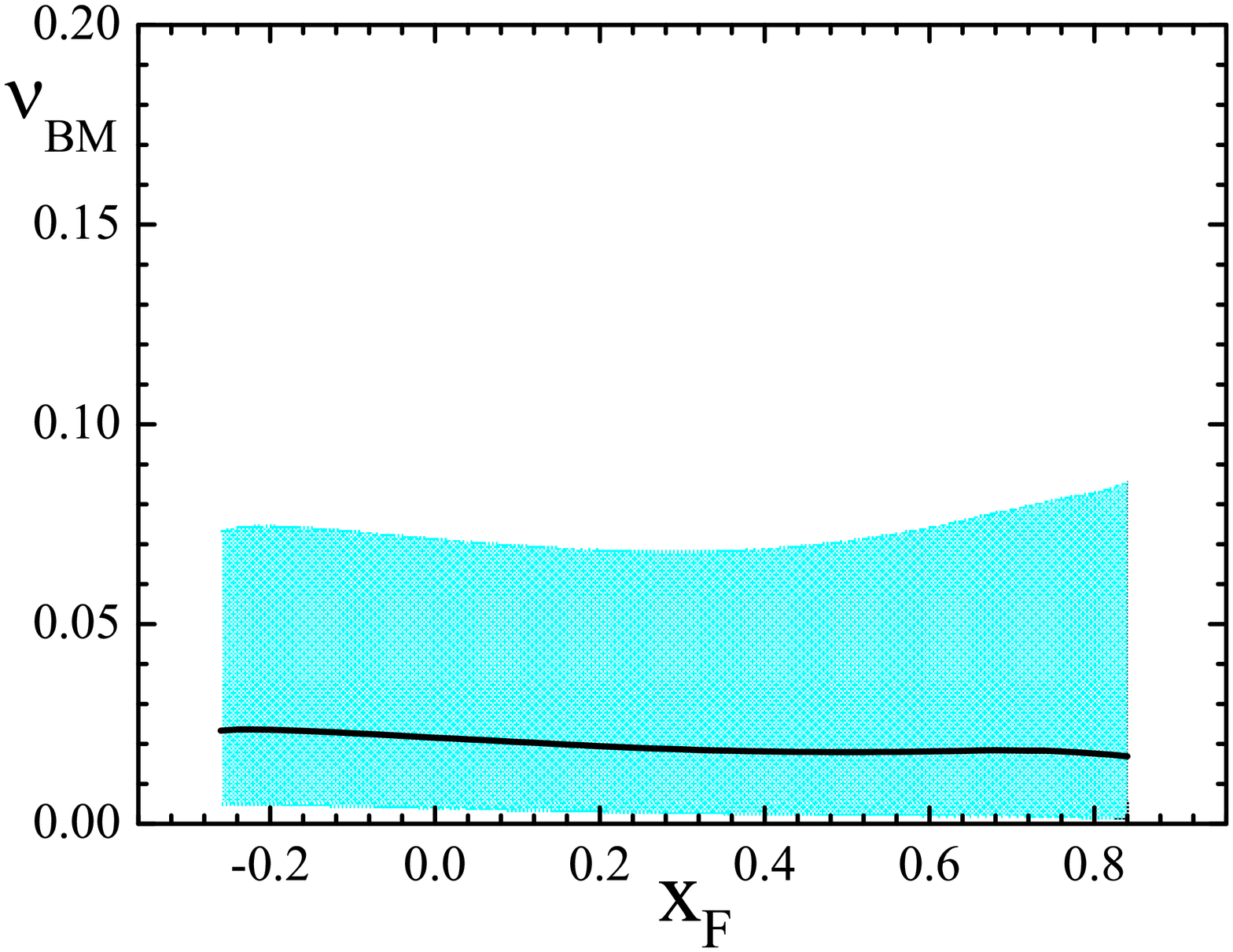}
  \includegraphics[width=0.48\columnwidth]{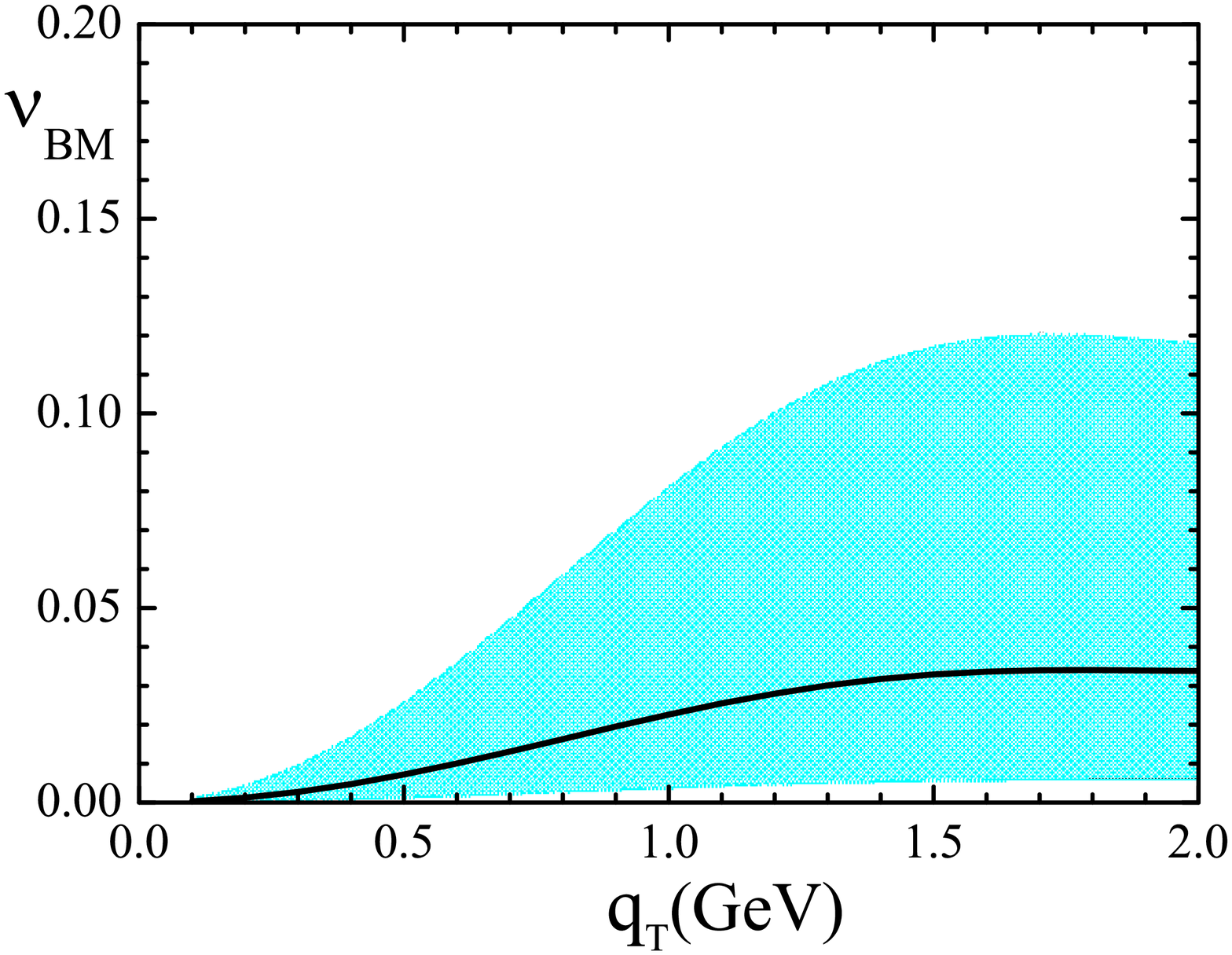}
  \caption{The $\cos2\phi$ azimuthal asymmetries $\nu_{BM}$ as the functions of $x_p$~(upper left), $x_\pi$~(upper right), $x_F$~(lower left) and $q_T$~(lower right) for the unpolarized $\pi p$ Drell-Yan process considering the TMD evolution at the kinematics of COMPASS. The shadow areas correspond to the uncertainty of the parameters in the parametrization of the Boer-Mulders function for proton in Ref.~\cite{Lu:2009ip}.}
  \label{fig:asy}
\end{figure}

The COMPASS Collaboration at CERN adopts a $\pi^-$ beam with $P_\pi=\ 190\ \mathrm{GeV}$ colliding on a $\mathrm{NH}_3$ target~\cite{Gautheron:2010wva,Aghasyan:2017jop}, which provides a great opportunity to explore the Boer-Mulders function of the pion. The kinematics of the $\pi^- p$ Drell-Yan process at COMPASS are as following
\begin{align}
&0.05<x_N<0.4,\quad 0.05<x_\pi<0.9, \quad
4.3\ \mathrm{GeV}<Q<8.5\ \mathrm{GeV},\quad s=357~\mathrm{GeV}^2,\quad -0.3<x_F<1.
\label{eq:cuts}
\end{align}

Using the expression of $\nu_{BM}$ in Eq.~(\ref{eq:nu}) as well as the denominator in Eqs.~(\ref{eq:denominator}) and the numerator in Eq.~(\ref{eq:numerator}), we calculate the $\cos2\phi$ Boer-Mulders asymmetry $\nu_\textrm{BM}$ as functions of $x_p,\ x_\pi,\ x_F$ and $q_T$.
In calculating the functions of $x_p-$, $x_\pi-$ and $x_F-$dependent asymmetries, the integration over the transverse momentum $q_T$ is performed over the region $0<q_T<2~\mathrm{GeV}$ to make the TMD factorization valid. The same choice has been made in Refs.~\cite{Sun:2013hua,Wang:2018pmx}.

We plot the results of $\nu_{BM}$ in Fig.~\ref{fig:asy}, in which the upper
panels show the asymmetries as functions of $x_p$~(left panel) and $x_\pi$ (right panel); and the lower panels depict the $x_F$-dependent~(left panel) and $q_T$-dependent (right panel) asymmetries, respectively.
The bands correspond to the uncertainty of the parametrization of the Boer-Mulders function of the proton~\cite{Lu:2009ip}.
We find from the plots that, in the TMD formalism, the $\cos2\phi$ azimuthal asymmetry in the unpolarized $\pi^-p$ Drell-Yan process contributed by the Boer-Mulders functions is around several percent.
Although the uncertainty is rather large, the asymmetry is firmly positive in the entire kinematical region.
The asymmetries as the functions of $x_p,\ x_\pi,\ x_F$ show slight dependence on the variables, while the $q_T$ dependent asymmetry shows increasing tendency along with the increasing $q_T$ in the small $q_T$ range where the TMD formalism is valid. Our results show that, precise measurements on the Boer-Mulders asymmetry $\nu_{BM}$ as functions of $x_p,\ x_\pi,\ x_F$ and $q_T$ can provide an opportunity to access the Boer-Mulders function of the pion, as well as to constrain the Boer-Mulders function of the proton.

\section{Conclusion}
\label{Sec.conclusion}

In this work, we have applied the formalism of the TMD factorization to study the $\cos2\phi$ azimuthal asymmetry contributed by the coupling of two Boer-Mulders functions, in the pion induced
unpolarized Drell-Yan process that is accessible at COMPASS. To do this, we have adopted the model results of the unpolarized distribution function $f_1$ and Boer-Mulders function of the pion meson calculated from the light-cone wavefunctions.
For the distribution functions of the proton target needed in the calculation,  we have applied available parametrizations.

We have also taken into account the TMD evolution of the pion and proton distribution functions. Specifically, we have utilized the nonperturbative Sudakov-like form factor of the pion TMD distributions extracted from the unpolarized $\pi N$ Drell-Yan data, while for the proton target, we have adopted a parametrization of the nonperturbative Sudakov form factor that can describe the experimental data of SIDIS, DY dilepton and W/Z boson production in $pp$ collisions. We have also assume that the Sudakov form factors for the Boer-Mulders function are the same as those for the unpolarized distributions $f_1$.

We have calculated the contribution of the Boer-Mulders functions to the $\cos 2\phi$ azimuthal asymmetry in the unpolarized $\pi^- p$ Drell-Yan process at the kinematics of COMPASS.
The predictions are presented as functions of the kinematical variables $x_p,\ x_\pi,\ x_F$ and $q_T$.
We find that, the double Boer-Mulders asymmetry in $\pi^- p$ Drell-Yan process calculated from the TMD evolution formalism is positive and is sizable, around several percent.
It shows that there is a great opportunity to access the $\cos 2\phi$ azimuthal asymmetry in the unpolarized $\pi^- p$ Drell-Yan process at
COMPASS and to obtain the information of the Boer-Mulders function of the pion meson. Furthermore, the calculation in this work will also shed light on the proton Boer-Mulders function since the previous extractions on it were mostly performed without TMD evolution.

\section{Acknowledgements}

Z.~L. Thanks Wen-Chen Chang for bringing our attention on the Boer-Mulders effect at COMPASS.
This work is partially supported by the National Natural Science
Foundation of China (Grants No.~11575043 and No.~11605297), by the Fundamental Research Funds for the Central Universities of China. X.~W. is supported by the Scientific Research Foundation of Graduate School of Southeast University (Grants No.~YBJJ1667).

\end{document}